# Accurate Prediction of Experimental Band Gaps from Large Language Model-Based Data Extraction

Samuel J. Yang, Shutong Li, Subhashini Venugopalan, Vahe Tshitoyan, Muratahan Aykol,
Amil Merchant, Ekin Dogus Cubuk, Gowoon Cheon[*]
Google

## Abstract

Machine learning is transforming materials discovery by providing rapid predictions of material properties, which enables large-scale screening for target materials. However, such models require training data. While automated data extraction from scientific literature has potential, current auto-generated datasets often lack sufficient accuracy and critical structural and processing details of materials that influence the properties. Using band gap as an example, we demonstrate Large language model (LLM)-prompt-based extraction yields an order of magnitude lower error rate. Combined with additional prompts to select a subset of experimentally measured properties from pure, single-crystalline bulk materials, this results in an automatically extracted dataset that's larger and more diverse than the largest existing human-curated database of experimental band gaps. Compared to the existing human-curated database, we show the model trained on our extracted database achieves a 19% reduction in the mean absolute error of predicted band gaps. Finally, we demonstrate that LLMs are able to train models predicting band gap on the extracted data, achieving an automated pipeline of data extraction to materials property prediction.

## 1 Introduction

Modern scientific literature encodes information on practically all technological materials, their structure, processing history and property measurements in the form of unstructured natural language buried in research articles. Collection and standardization of such information at scale would transform how materials research is conducted. Human curated tables or handbooks of material properties are inherently limited in size and scope. They often lack critical structural and processing details that influence the properties, which leads to poor predictability of material properties. Rule-based approaches have been widely explored to automate the extraction of larger datasets from the literature,[25, 18, 8] but they also often fail to capture such critical information, and are challenging to benchmark or execute at high accuracy. Machine learning models trained on largely-available computational data,[16, 23, 6] such as density functional theory (DFT) computed band gaps, do not generalize to the experimental measurements either [1]. This is not only due to inherent limitations in the theory itself (e.g. underestimation of band gaps), but again due to the uncertainty in experimental data due to lack of information essential for predictions.

With the recent advances in natural language processing, we observe a significant improvement in the capabilities of Large Language Models (LLMs). With LLMs, prompt-based extraction from the scientific literature is emerging as a new extraction method that may achieve higher extraction accuracy at a lower amount of human effort. Early LLM-based property extraction studies utilized fine-tuning [9], but there have been a series of papers using prompt engineering with success, such as extraction of critical cooling rates of metallic glasses, yield strengths of high-entropy alloys, emission wavelengths of phosphors or synthesis parameters metal organic frameworks. [30, 31, 21, 20]

---

[*]corresponding authors: gowoon@google.com

In this work, we demonstrate that databases created from LLM-assisted, prompt-based extraction lead to state-of-the-art performance on predicting experimentally measured properties. We focus on the extraction of experimental band gaps because they involve the main extraction challenges outlined above: i.e. it is a property strongly affected by material morphology (e.g. film thickness or particle size) and doping levels. In fact, it is common practice to tune the band gap of a material by changing the nanoparticle size[29] or doping[22], and the exclusion of these effects leads to highly noisy datasets upon extraction. Moreover, it cannot be accurately predicted by the level of DFT used in high-throughput databases of materials, which are widely used for training machine learning models predicting band gap. To remove errors from DFT calculations in models predicting band gap, we exclude numerically calculated values of band gaps in our data extraction process.

Even with advanced data extraction methods, the challenges with data extraction as well as uncertainty in experimental data result in noise in any automatically extracted dataset. We demonstrate that using machine learning techniques that are robust to noise, such as huber loss[15, 13] and ensembling, improve the performance of the model trained on extracted data.

Our contributions are: (i) benchmarking the quality of extracted databases directly on the accuracy of models trained on the databases (ii) a database of experimentally measured band gaps of pure single-crystalline bulk materials extracted by prompt-based methods (iii) graph neural network models trained on the extracted dataset, which achieve state-of-the-art results on predicting experimentally measured band gaps (iv) automated training of band gap prediction models using LLMs, completing the automation from data extraction to materials property prediction.

## 2 Existing Datasets

There are three existing datasets that are most relevant to our work, and we use these to compare the value of our dataset and property prediction models.

**Dong&Cole[8], Auto-generated database of band gaps**: ChemDataExtractor[25, 18] is a leading software for extracting chemical information from scientific papers, and is widely used in the scientific community for materials and chemistry data extraction[8, 7, 14, 24, 3]. Not only is it widely used, many materials and chemistry NLP methods use ChemDataExtractor's tokenization as well[27, 26, 17]. Ref.[8] applied it to sentences from 128,776 journal articles to extract band gaps and reports 84% precision and 65% recall. As is the standard for data extraction, the precision and recall are evaluated on whether the extracted values match the original sentence. The data points used to compute these metrics contain materials of varying morphology and composition - for example, it contains over 500 entries for GaN with standard deviation 1.17 eV. Hence, using just the precision and recall does not fully capture the utility of the extracted dataset for predicting band gaps of materials.

The dataset released from this work include both the extracted band gaps and the original sentences where the values were extracted. We use the sentences in this dataset as the source of text for our LLM-based extraction. In the rest of the paper, we refer to the original ChemDataExtractor-extracted values from this dataset as *Dong&Cole*, and our LLM-based extractions from the same set of sentences as *Our extractions*. For chemical compositions that have multiple extracted band gap values, we take the median to have a single value per chemical composition.

**Materials Project**: The Materials Project database is one of the most widely used datasets for training machine learning models for materials. Though the error of DFT-computed band gaps in a high-throughput setting may be large[1], we include this dataset as it is a popular benchmark dataset for state-of-the-art property prediction models including band gap. We use the snapshot of the Materials Project as of June 1, 2018, which was released with the MEGNet paper[5]. For chemical compositions with multiple phases, we take the band gap value of the structure with the lowest energy above the hull.

**matbench_expt_gap**: This is the database of experimentally measured band gap values from the literature[32], which is used as a benchmark dataset for evaluating models for band gap prediction in Matbench[10]. We use the version of the dataset provided in Matbench, which has been edited to deduplicate values for the same chemical composition. However, only the non-zero band gap values in this dataset come from the literature: the authors note in the original paper[32] that "The experimental band gap values were extracted from literature as referenced and for metals they were extracted from Materials Project database." We checked the subset of non-metals in matbench_expt_gap that also exists in Materials Project, and found that 17% of the non-metals in matbench_expt_gap are labelled as metals in the Materials Project. Finding this, we remove all metals (zero band gap materials) from our evaluation to ensure the comparison is done only on experimental band gap values from the literature without noise from DFT.

## 3 Data extraction

**Data extraction and filtering** We use the sentences in the dataset from Dong&Cole as the source of text. Inspired by [30], our approach uses a series of zero-shot [4] prompting of LLMs specifically tailored to identify and extract materials and their properties. We use GPT-4[19] as the LLM. First, for each material whose band gap value is described in a sentence, we extract the material, property name, property value and unit, as well as any descriptors for the material (e.g. "nanoparticle") or property (e.g. "DFT-computed"). For each extracted entry, we run four follow-up prompts to ask: 1) Is the extracted property actually a band gap value?(as opposed to, for example, increase in the band gap, or +U correction in eV units) 2) is the extracted material a pure (non-doped) bulk single-crystalline material? 3) What is the chemical formula of the extracted material? 4) Is the band gap numerically calculated?

All prompts that were used in this work can be found in the Supporting Information section. After running the prompts, we filter the data for experimentally measured electronic band gaps of pure single-crystalline bulk materials. We make this choice so that our data reflects the intrinsic properties of materials, rather than properties engineered through changes in morphology or composition. Moreover, this provides a direct point of comparison with numerically calculated values. The following inclusion criteria were applied: prompt responses have correct formatting; there is no evidence the extracted property was numerically computed; there is no evidence the extracted entry is not for a pure single-crystalline bulk material; the extracted units and property names are consistent with band gap; band gap value is between 0 and 20 eV.

The reason for manually excluding extracted values that are over 20 eV, instead of having a full prompt-based pipeline, is because of errors from the original source text. For example, the sentence from [28], *As illustrated in Fig. 3a, the band gap of the Zn16O16 system was 3400 eV, which is consistent with that in a previous study [26].*, has the typographical error *3400 eV* instead of *3.400 eV*. Even in the limit of achieving perfect precision for extracting values that match the original text, the errors from the original text still persist. This yields 39391 extractions for 2733 distinct compositions. For the 1336 compositions that have multiple extractions per material, we take the median of the extracted band gap values.

**Verification** To verify the quality of our extractions, we conducted several analyses. First we manually annotated 100 randomly selected sentences and evaluated the precision and recall across our various prompts. For entity extraction, we assess whether the identified property, value and units matches the material, and that all are contained in the source sentence, as done in ChemDataExtractor[25, 18]. As seen in Table 1, we achieve an order of magnitude lower error rate, with a 99% precision compared with 81% for ChemDataExtractor.

Next, we evaluated our prompts for classifying whether extracted material properties are numerically computed or from materials that are not pure single-crystalline materials in the bulk form. Where the sentence provides positive evidence of either, we exclude the extraction from our dataset; in cases of negative or lack of evidence, we include the extraction. We evaluate these prompts in this binary way, with the precision and recall reported in Table 1.

Next, we compare the our extracted dataset with the values extracted in Dong&Cole. Among 4871 compositions in the Dong&Cole dataset, 2733 compositions are contained in our dataset. Only 52% of the compositions in both datasets have band gap labels within 0.1 eV.

## 4 Evaluation

Our evaluation compares the value of the extracted data in training performant experimental band-gap prediction models. As noted above in the description for matbench_expt_gap , we remove all metals from all 4 datasets (Dong&Cole, our extractions, matbench_expt_gap , Materials Project) before we perform the evaluations. We train a band gap property prediction model on each of the datasets in Sec. 2, excluding materials in the test set described below. Another experiment we run is to train a model on the superset of our extractions and matbench_expt_gap . As this is the largest dataset of experimental band gaps to date, we expect the predictions from models trained on this dataset to be the most accurate and useful for predicting band gaps of new compounds.

|  | (material, property, value, units) extraction | | | property_is_numerical | | material_is_not_crystalline | |
| --- | --- | --- | --- | --- | --- | --- | --- |
|  | Dong&Cole | prompt 1* | prompt 1* & 2** | Dong&Cole | Ours | Dong&Cole | Ours |
| precision | 81% (81/100) | 97% (97/100) | 99% (99/100) | N/A | 70% (7/10) | N/A | 79% (38/48) |
| recall | 72% (67/93) | 96% (89/93) | 92% (86/93) | N/A | 89% (8/9) | N/A | 65% (41/63) |

Table 1: Evaluating extractions from Dong&Cole, which uses ChemDataExtractor, and our proposed prompt-based extraction on 100 randomly selected human-annotated sentences encompassing 140 property extractions. Extraction precision represents fraction of sentences where no incorrect extractions were obtained; extraction recall represents sentences where all extractions were found (only 93 of 100 sentences had at least one valid extraction, per human annotation). For classification (if the property is numerically computed and not experimentally determined; if the material is not pure bulk single crystalline), precision and recall denote fraction of extractions.
*: prompt 1 extracts from a sentence all material(s), property name(s), property value(s) and unit(s), as well as any descriptor(s).
** prompt 2 confirms the extracted property is a valid band gap. Both the prompts can be found in the Supporting Information.

We evaluate the models in two different ways. First, we report the 5-fold cross validation error of models trained on each dataset. However, the datasets have different standard deviations as well as different distribution in the chemical composition space. A raw numerical comparison of cross validation errors is unlikely to be predictive of how well the models will perform on predicting the band gaps of new materials.

Hence, we perform a second evaluation on a shared hold-out test set, consisting of 210 materials that are common across all datasets mentioned above. The fact that these materials appear in all datasets indicate research interest, so these materials are more likely to resemble the distribution of materials that materials scientists are interested in studying.

**Graph neural network model used for predicting band gaps** We train GNN models to predict experimental band gap, following the message passing formulation [2, 11] and implemented using Jraph [12]. We use the "matscholar_el" [27] featurization for node features, and edges are formed between all pairs of nodes in the graph. The models update latent node features via stages of message passing, where neighbor information is collected via normalized sums over edges and representations are updated via shallow MLPs [11]. After several steps of message passing, a linear readout layer is applied to the global state to compute a prediction of the energy. The GNN used an embedding size of 64 and two hidden layers of size 164 and 64. The model is trained for 1000 epochs. To improve the predictions, we use an ensemble by taking the mean prediction across 10 models with different random initializations.

To evaluate how good this model architecture is compared to other models, we evaluated our model on the matbench leaderboard [10] for predicting experimental band gaps. Our ensembled model architecture achieves 0.343 eV mean absolute error, which is on par with the best submission in the matbench leaderboard (0.331 eV for Ax/SAASBO CrabNet v1.2.7).

**The choice of loss function** Though our prompt-based extraction substantially improves the precision of the extracted data, we still find that in many cases, the extracted band gaps do not have a single value for each material. The main reason for this is because our data source is single-sentence excerpts from papers. For example, consider a paper describing the fabrication of a ZnO thin film in the methods section, and mentions the measured band gap of the ZnO sample in the Results section. If the sentence containing the band gap value does not refer to ZnO as a thin film, our extraction pipeline does not have enough information to decide whether this is a bulk sample of ZnO. As mentioned in the Data extraction and filtering section, errors from the source sentences may also introduce noise in the dataset, as would LLM errors in the extraction pipeline. As our data is inherently noisy, we use the Huber loss, which is known to penalize outliers in the data during training. We compare the results using Huber loss and the standard L2 loss in the Supporting Information.

## 5 Results and Discussion

**5-fold cross validation** Table 2 presents our main results. While the cross validation errors on the matbench_expt_gap leaderboard are 0.3 eV, 53% of the materials in the leaderboard are metals.

|  | Size | 5-Fold | Test MAE | |
|---|---|---|---|---|
| Dataset | (# materials) | CV | (our labels) | (matbench labels) |
| Dong&Cole | 4871 | 0.798 | 0.700 | 0.862 |
| Ours | 2733 | 0.671 | 0.592 | 0.714 |
| matbench_expt_gap | 2152 | 0.405 | 0.798 | 0.772 |
| Ours+matbench_expt_gap | 4495 | 0.550 | 0.607 | 0.626 |
| Materials Project | 30834 | 0.434 | 1.105 | 1.205 |

Table 2: We report Mean Absolute Error (MAE) in eV, comparing band gap prediction models trained on different datasets

Though cross validation errors on matbench_expt_gap in our table seem a lot worse than the matbench_expt_gap leaderboard, this is expected from removing metals. (As noted in the Evaluation section, our architecture achieves close to state-of-the-art performance on the matbench_expt_gap leaderboard when metals are not removed.)

Compare the original Dong&Cole dataset to our extractions, which applied additional filtering on the same source sentences. While the Dong&Cole dataset is 60% larger than ours, the cross validation error is 16% lower in our dataset. This indicates that our filtering method was effective in filtering out noise in the dataset.

While the cross validation error of our dataset is larger than that of matbench_expt_gap, the standard deviation of our dataset is 1.78 eV, whereas it is 1.47 eV for matbench_expt_gap, so the difference in cross validation error is actually smaller than the difference in standard deviation. We plot the distribution in the Supporting Information.

**Shared hold-out test set** We demonstrate that the model trained on our extractions outperform the model trained on matbench_expt_gap, and the difference is even larger for the model trained on both datasets.

Note that for the shared test set, some materials have different band gap labels in our dataset and in matbench_expt_gap. matbench_expt_gap labels are human curated and are used for benchmarking experimental band gaps, so we focus on the evaluation using matbench_expt_gap as ground truth. This is a more difficult evaluation on the model trained on our dataset, compared to the model trained on matbench_expt_gap. However, we see that the model trained on our dataset outperforms the model trained on matbench_expt_gap, even when using matbench_expt_gap labels as ground truth. This indicates that that our extracted dataset is better than matbench_expt_gap as a training set for band gap prediction models. (We did not compute the MAE using Materials Project labels, as they are based on DFT computations)

Most importantly, the combined dataset of our extractions and matbench_expt_gap reduces the band gap prediction error by 19% on the test set. This demonstrates that our extracted dataset is complementary to the existing matbench_expt_gap, and our extraction method is effective for improving the prediction of experimental band gaps.

Finally, we include Materials Project band gaps in the comparison. While Materials Project band gaps are widely used for training and benchmarking band gap prediction models, we see that training on the Materials Project data is not effective for predicting experimental band gaps.

**Using LLMs to train models on extracted data** With the extracted data, the coding ability of the LLMs also enables us to simply prompt the LLM to train property prediction models from the dataset. We show prompts that use simple natural language instructions to make the LLM load and manipulate the data. The LLM then goes on to generate and execute the code to perform the manipulations and also trains and compares logistic regression, gradient boosted trees and random forest models for band gap prediction. We share a colab notebook demonstrating this at https://github.com/google-research/google-research/tree/master/matsci.

## 6 Conclusion

In this work, we demonstrate that we can extract and construct a high quality material properties dataset from literature with the assistance of LLMs. Taking experimental band gap of pure single-crystalline bulk materials as an example property, we construct a dataset of 2733 materials. We benchmark the quality of our dataset comparing with existing datasets and show that our dataset is useful for training property prediction models that achieve state-of-the-art performance on predicting experimental band gaps. We also demonstrate a path forward for automated modeling of material properties from extracted data.

# Supporting Information

### Distribution of our dataset and matbench_expt_gap dataset

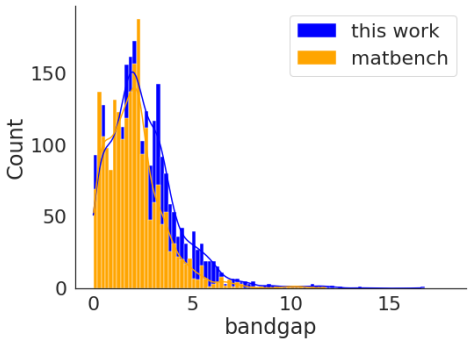

The distribution of our dataset and matbench_expt_gap dataset (excluding metals). Our dataset is both larger and has a more dispersed distribution compared with matbench_expt_gap dataset: the mean and standard deviation of our dataset are 2.58 and 1.87 eV, whereas for matbench_expt_gap they are 2.09 and 1.47eV.

### Dataset sizes with metals

Table 3 presents the sizes of the different datasets with metals.

|  | total number of data entries | number of unique materials with metals | number of unique materials without metals |
|---|---|---|---|
| Dong&Cole | 54053 | 4873 | 4871 |
| Ours | 39391 | 2733 | 2733 |
| Matbench | 4604 | 4604 | 2152 |

Table 3: The number of raw data entries allowing multiple band gap labels for each material, and the number of unique materials including metals. In the experiments in the main text, only the unique materials that are non-metals are used.

### Results using the L2 loss

Table 4 presents the results when using L2 loss when tuning the GNN models.

| Dataset | Size (# materials) | 5-Fold CV | Test MAE (our labels) | Test MAE (matbench labels) |
|---|---|---|---|---|
| Dong&Cole | 4871 | 0.794 | 0.680 | 0.844 |
| Ours | 2733 | 0.675 | 0.632 | 0.741 |
| matbench_expt_gap | 2152 | 0.405 | 0.775 | 0.757 |
| Ours+matbench_expt_gap | 4495 | 0.552 | 0.597 | 0.645 |
| Materials Project | 30834 | 0.435 | 1.109 | 1.213 |

Table 4: Mean absolute errors computed using the L2 loss

### Results using stratified sampling

Another option we considered for handling multiple different labels in the extracted data is to train different ensembles on a different subset of extracted labels. For each ensemble, we sampled 5 extracted labels for each material at random. Results for this stratified sampling during training are presented in Table 5.

| Dataset | 5-Fold CV | 5-Fold CV, SS | Test MAE (matbench labels) | (matbench labels), SS |
| --- | --- | --- | --- | --- |
| Ours | 0.671 | 0.676 | 0.714 | 0.749 |
| Ours+matbench_expt_gap | 0.550 | 0.544 | 0.626 | 0.651 |

Table 5: Mean absolute errors computed using the Huber loss and stratified sampling (SS). The columns with "SS" used 5 random samples of extracted labels for each material during training. The MAE was evaluated using the median value for each material.

**Prompts**

**prompt 1**: Extract properties from a passage. "[[material_property]]" is "band gap".

```
For each material whos [[material_property]] property and value is described in the
    following passage from a scientific paper, extract the following fields:

1. material: material chemical formula. required.
1. property: material property for the named material. required.
1. units: property units
1. value: property value. required. must be a quantitative integer or float, not
    qualitative.
1. material descriptors: material descriptors or 'None' if none are present
1. property descriptors: property descriptors or 'None' if none are present
1. is_evidence_material_is_not_pure_bulk_crystalline_material: 'evidence_is_not' or
    'evidence_is' or 'no_evidence_either_way', whether the material is a pure (non-
    doped) bulk crystalline material?

If any required field is absent, exclude that result entirely. Do not mark any
    fields as 'unknown'.

Example output:
[{'material': 'Si',
  'property': 'melting point',
  'unit': 'C',
  'value': '300',
  'material descriptors': 'crystalline',
  'property descriptors': 'sea level',
  'is_evidence_material_is_not_pure_bulk_crystalline_material': 'evidence_is'},
  ...]

Passage from a scientific paper:
[[passage]]

<<reply>>
```

**prompt 2**: Confirm an extracted property is valid. For this and following prompts, the "[[extraction]]" is the extraction's material, property, units, value, material descriptors and property descriptors, JSON-formatted.

```
For the information extracted from the following passage from a scientific paper, is
    the following material property the actual or predicted bandgap value for the
    specific material?

Example input:
1. Extraction: {"formula":"AlCuS2","bandgap":1.0,"material descriptors":"crystal","
    property descriptors":"decreased"}
   Passage from a scientific paper: In most of crystals from this group, including
       CuAlS2, the bandgap is decreased by 1 eV as compared to the ZnSe analog, and
       the spinorbit splitting decreases from 0.45 to 0.23 eV.
2. Extraction: {"formula":"CeO2","bandgap":7.0,"material descriptors":"None","
    property descriptors":"None"}
   Passage from a scientific paper:A U value of 7.0 eV was used to reproduce the
       correct band gap for CeO2.
```

```
Example output:
[{'is_actual_bandgap_value': 'no',
  'reason': 'the absolute bandgap is not mentioned in this passage.'},
  {'is_actual_bandgap_value': 'no',
  'reason': 'U value is used as a parameter, not the actual bandgap value.'},
  ...]

Extraction:
[[extraction]]

Passage from a scientific paper:
[[passage]]

<<reply>>
```

**prompt 3**: Check if extracted property is numerically calculated.

```
For the band gap information extracted from the following passage from a scientific
    paper, \
I want to determine if the band gap was calculated numerically or from analyzing
    experimental measurements. \
An experimental measurement of band gaps usually involves absorption-based
    spectroscopic techniques, \
such as UV-vis, photoluminiscence, diffuse-reflectance, angle-resolved photoemission
     spectroscopy. \
Then the band gap is calculated from the experimental results, using extrapolation,
    \
Kubelka-Munk transformation, Tauc equation, absorption spectrum fitting, Beer-
    Lambert's law etc.
If the band gap came from numerical calculations, it will mention that it was
    calculated through simulations, theoretical calculations, \
or methods such as density functional theory(DFT), many-body calculations or tight
    binding; \
or the sentence may contain keywords that are related to DFT, such as functionals
    PBE, GW, G0W0, SCAN, B3LYP, HSE, mBJ, PBESol \
or other keywords such as Hubbard U, +U calculations, Hartree-Fock, exchange, plane
    wave basis, gaussian basis, density of states etc. \
In this case, the band gap definitely does not come from experimental measurements.
Is the given extraction from the sentence a value measured experimentally? Which
    method was used?
Example outputs:
{'answer_is_experimental_method': 'yes', 'method':'Diffuse-reflectance measurement',
    \
'reason': 'The sentence mentions Kubelka-Munk relation, which is used for converting
     the diffuse reflectance spectra to pseudo-absorption spectra'}
{'answer_is_experimental_method': 'yes', 'method':'unknown experimental method', \
'reason': 'The sentence mentions DFT calculation result of 1.3 eV, but also says
    that it matches the experimental value of 1.2 eV; \
the given extraction was 1.2 eV, which is the experimental value.'}
{'answer_is_experimental_method': 'no', 'method':'DFT', \
'reason': 'The sentence mentions that U value of 10 eV was used, which is the
    Hubbard term for DFT calculations of materials with d or f electrons.'}
{'answer_is_experimental_method': 'no', 'method':'DFT', \
'reason': 'The sentence mentions functional, which is used in DFT calculations.'}
{'answer_is_experimental_method': 'unknown', 'method':'unknown', \
'reason': 'The sentence does not mention whether the band gap was measured or
    calculated, or include any details about the measurement (or calculation)
    method.'}

Extraction:
[[extraction]]

Passage from a scientific paper:
[[passage]]
```

```
<<reply>>
```

**prompt 4**: Check if extracted material is not crystalline

```
For the material property information extracted from the following passage from a
    scientific paper, is the extracted material a pure (non-doped) bulk single
    crystalline material?

Example output:
[{'answer_yes_no_or_not_enough_information': 'yes',
   'reason': 'reason'},
   ...]

Extraction:
[[extraction]]

Passage from a scientific paper:
[[passage]]

<<reply>>
```

**prompt 5**: Check if extracted material formula is valid.

```
    What is the chemical formula of material [[material]]? Respond with only a
        chemical formula.

<<reply>>
```